# COMBINE: a novel drug discovery platform designed to capture insight and experience of users

Sung Jin Cho*

*CimplSoft, Thousand Oaks, CA  91320.*

*Corresponding author phone: (805) 807-3913; e-mail: sjcho@cimplsoft.com.

**ABSTRACT**

The insight and experience gained by a researcher are often lost because the current productive and analytics software are inherently data-centric, disconnected, and scattered. The connected nature of insight and experience can be captured if the applications themselves are connected. How connected applications concept is implemented in **CO**nstruct che**M**ical and **BI**ological **NE**twork (**COMBINE**), a novel user-centric drug discovery platform, is described. Using publicly available data, how COMBINE users capture insight and experience is explained, and how COMBINE users perform data organization, data sharing, data analysis, and data visualization is illustrated.

**INTRODUCTION**

Drug discovery is a complicated, capital-intensive, and lengthy process requiring expertise from multiple scientific disciplines. Because the process is multidisciplinary in nature, data sets generated and used by drug discovery researchers are as diverse as the fields they are in. In addition, advances in instrumentation and information technology led to unprecedented availability of scientific and meta data that are complex and highly interconnected. Many attempts were made to leverage them to expedite the drug discovery processes.[1,2,3,4,5,6,7,8,9,10,11,12,13] And recent successes in applying deep learning in computer vision, natural language processing, speech recognition and generation, self-driving cars, games, etc. led to increase interest in applying machine learning to solve drug discovery problems.[14,15,16,17]

Past breakthroughs in artificial intelligence (AI) closely follow the availability of digitized databases,[18,19] and it is encouraging to see that many publicly available chemical and biomedical databases exist and are growing both in size and diversity (Table 1).[20] Many commercial and open source tools[21,22,23,24,25] that utilize them are available to drug discovery scientists and are invaluable in data processing, analysis, and visualization. Many AI experts are also actively exploring ways to apply state-of-art machine learning methods to expedite the drug discovery process.[15,16,17,26] Undoubtedly, drug discovery researchers will benefit from the current effort and the availability of yet to be created machine learning approaches.

Common characteristics, found among these tools, are that they are inherently data-centric, disconnected, and scattered. For example, performing a routine medicinal chemistry task of performing similarity and substructure searches, followed by looking for pharmacology and protein structure data, requires juggling multiple tools and resources. Saving the search results is tedious, and often the entire process needs to be repeated multiple times using different starting or sub structures to explore design hypotheses (Figure 1). Remarkably, despite the obvious inconvenience frequently experienced by researchers, there are no tools to address this balancing act of exploring hypotheses and keeping track of what has been done. The insight and experience gained by researchers are found in their activity history, which is a valuable information not utilized fully by the current productive and analytics software.

In this paper, **CO**nstruct che**M**ical and **BI**ological **NE**twork (**COMBINE**), a novel user-centric drug discovery platform is described. The connected nature of insight and experience can be captured if the applications themselves are connected, and how this "connected applications" concept is implemented in COMBINE is described. How data organization, data sharing, data analysis, and data visualization can be performed differently compared to traditional approaches are illustrated using publicly available data.

**RESULTS and DISCUSSIONS**

**COMBINE**

COMBINE is a standalone software written using c/c++ and Qt[27], a c++ cross-platform application framework.  COMBINE is a data source agnostic and is extremely flexible in dealing with diverse types of data.  Web APIs are used to access various resources described in Table 1.  When necessary, web pages can be parsed, a socket connection to a server can be made to communicate directly over a specified port, or random accessible binary data files can be generated and used to avoid overhead associated with relational databases.  Every application in COMBINE, called app node, has a data table and a hash table.  Data stored in app node's tables are used to create an interactive visualization.  Figure 2 shows the screen shot of different app nodes found in COMBINE.

Connected app nodes in COMBINE are represented as directed and undirected graphs, called a knowledge network where edges describe relationships between app nodes, between a data point in an app node to another app node, and between a data point in an app node to a data point in another app node (Figure 3).  Each edge is added interactively either by simply double clicking the data specified by a mouse point or by selecting one of app node specific menus.  The double-click action or selecting a menu also instantiates an appropriate app node, and this interaction-triggered instantiation is how COMBINE tracks users' activities and allows users to focus on the current activity without losing the context.  App nodes and edges connecting them can be annotated to describe them further if the default visualization is not intuitive enough, and tool tips are used to provide help messages.

**Dependency**

OpenSSL[28] is used to provide secure communications between COMBINE and different web servers.  The FFmpeg library[29] is used to extract frames out of a video file, and the RDKit library[24] is used to perform various cheminformatic tasks. In addition, command line programs from Open Babel[30], the Indigo toolkit[31], and the OFFIS DICOM toolkit[32] are used to convert file formats.

**Public Databases & APIs**

Many public databases and APIs are incorporated into COMBINE (Table 1), and they can be accessed by creating an app node containing the data or interacting with the app node. An example of how different resources can be stringed together seamlessly to address a typical medicinal chemist's workflow is described in Figure 1. The ChEMBL[40] database is a database containing structures and annotations of over 1.7 million compounds, and it is a database frequently used to explore a design hypothesis. A similarity searching can be performed against it, or different chemotypes can be used to perform substructure searches. Once interesting ChEMBL hits are found, Open PHACTS APIs[41] can be used to query various pharmacological data as well as patent data found in the SureChEMBL[42] database. Many different structure identifiers, representing the same structure, are possible, and the UniChem[43] database provides a way to cross-reference against 33 data sources currently available. One of the pharmacological data linked to the ChEMBL compound is the molecular target information, which is typically determined by examining its activity value and the type of biochemical assays it is active against. The default pChEMBL[33] value of 6 is used to make the determination unless noted otherwise. The protein sequence and functional information of the molecular target are found in the UniProt[44] database. If publicly known, the 3D structure of the protein target can be obtained from the PDB[45] database. Like small molecule cross-references provided by the UniChem[38] database, the BioDBnet[46] database provides molecular target cross-references and is used to convert from gene ID to UniProt ID. Protein and genetic interactions, chemical interactions, and post translational modifications data are stored in the BioGRID[47] database.

**Data Organization and Data Sharing**

Like tables found in a relational database management system, each app node stores and manipulates the data in tabular form, and the data found in each cell in the table can be of diverse data types. Since app nodes are connected, carrying all the information all the time is not necessary. Each app node only needs to keep track of the parent app node where the current app node was generated from and extract additional information when necessary. Since an edge can also originate from a data point, its location must be tracked as well. The data stored in each app node can be also transferred from external data sources via web services. Stored data can be permanently associated with each app node or, when more convenient, retrieved in real time when a saved knowledge network is loaded. In this case, the knowledge network contains a list of URLs describing how different data can be retrieved. A PDB node, for example, retrieves the image of the selected protein when it is loaded.

A knowledge network is generated as users interact with each app node, and edges (either lines or arrows) describe how different activities are connected (Figure 3). This simple feature allows a user to visualize the history of how data was transformed and/or analyzed. A typical workflow of medicinal chemists involves finding similar compounds to the compound they are working on. From hits they found, they search for any associated pharmacology data. If a connection to a protein target can be made, other targets, pathways they are in, and diseases they are implicated in are all the information medicinal chemists are interested in knowing. This workflow is illustrated in Figure 4a. What many standalone or browser enabled applications typically allow users to do is to save the result or bookmark the relevant web pages. How users arrive at that position must be described separately or, most of times, is lost. The knowledge network in COMBINE captures this history automatically, so that users know exactly how they arrived at the result. Sharing the history improves productivity because seeing what has been done allows users to minimize the chance of repeating the work, and they can plan things more efficiently. It also promotes a collaboration because other users can follow the history to better understand what has been done, repeat the work if desired, and build upon the previous work easily. Continuing the example given earlier, based on literature hits, the medicinal chemist can perform the structure activity relationship analysis, design a new compound, and search against a patent database (Figure 4a). A biologist can look for other off target activities of the designed compound and formulate novel therapeutic uses (Figure 4b). An in vivo pharmacologist can link imaging data to correlate in vitro assay data or add a short video to show the behavior of an animal.

One common activity most researchers perform is to read scientific papers and try to apply what they learn in their own research. Writing a research paper to describe positive findings is a typical result. However, negative findings and other exploratory activities are not captured properly. COMBINE allows users to capture these activities by linking datasets, performing analyses, and visualizing results. Figure 4b illustrates different user activities stemming from a medicinal chemistry paper. Target information is added to describe the biology of the capsaicin receptor, and the structure-activity relationship table is generated using compounds described in the paper. Annotating a research paper with a dataset which a user can interact in real time is an extremely powerful way to understand and apply their findings.

**Data Analysis and Data Visualization**

Data analysis can be performed locally or remotely. This decision is largely based on the accessibility of remotely available tools, but an emphasis is given to a design that will improve the user experience. Generating fragments, calculating properties, performing various clustering, etc. are examples of tasks that are performed locally because they can be processed relatively fast. The result generated from executing each task is encapsulated in another app node and connected automatically to the parent app node. As described previously, this frees up valuable research time, and users can focus more on design and exploring hypotheses. The default visualization produced by each app node is designed carefully to be intuitive and concise.

An app node can also be configured to act as user interface (UI) to control applications installed remotely. Figure 5 illustrates a simple app node containing 3 buttons to retrieve three types of data from Firebrowse[34] using their Web APIs. The first button is to access 38 different cohorts, the second button is to access mRNAseq expression profiles found in different cohorts, and the third button is to access the copy number variations of participants found in different cohorts. For example, double clicking the Cohort button creates the "firebrowse cohorts" app node, retrieves the list of 38 cohorts and their description, populate the app node's data table, and visualize the list as interactive buttons. Double clicking the button containing adrenocortical carcinoma retrieves participants belong to the cohort, and a custom plot, combining box and scatter plots, are created to visualize downloaded data. Each additional query action, performed by double clicking a data point or a button, adds a connection to the knowledge network. Once results are stringed together this way, other data, like infographics describing the disease or the structure of a drug, can be easily added to describe the network further. Such app nodes are also good branching points to introduce different data.

A more sophisticated UI example is Cas9 gRNA tool (Figure 6). This UI collects an input DNA sequence and runs server-side programs, including Bowtie[35], an ultrafast short sequence alignment program. The web version of the application is found at https://cheminformatic.com/grna/index.php. It outputs unique 23mers, corresponding target sequences (reverse complements if the direction shown is "-"), GC content, the number of off-targets for 1bp and 2bp mismatches, and the sequences of off-targets. A "grna tool" app node, an interactive DNA sequence viewer, is used to specify locations of target sequences, and the tabular form of the output which is stored in the data table can be displayed.

The Details-on-demand[36] (DoD) visualization approach is heavily utilized at the network and app node level. It is a technique used to free up the computer resource when displaying details is not necessary. The network level DoD visualization is activated when no app node has been selected, and zoom levels and the area of network currently displayed in the viewing window are used to determine how the computer resource should be used. The part of the network outside the viewing window is excluded from drawing, and app nodes that are too small to see are converted to static images. Static images are converted to interactive visualizations only when sufficiently zoomed in so that individual components can be identified. This technique allows users to create a knowledge network containing a large number of app nodes and to navigate easily regardless of the size of the network. The preview panel is used to move quickly to the desired area (Figure 7). This feature is especially useful when there are many disconnected knowledge networks.

Once an app node is selected, the app node level DoD visualization is activated. Figure 8 shows how the app node level DoD visualization technique is used to visualize a matched molecular pair (MMP) network[37] created using 782,524 ChEMBL compounds with 14,680,477 edges which represent MMP pairs. The minimum spanning tree algorithm is used to reduce the number of edges that need to be drawn. Once a layout was generated using LGL[38], seven different size images (256x256, 512x512, 1024x1024, 2048x2048, 4056x4056, 8192x8192, and 16384x16384 pixels) were generated for each zoom level, and 256x256 pixel size tiles were cut out of each image. This cut out process produces 1, 4, 16, 64, 256, 1024, and 4096 tiles for zoom level 0, 1, 2, 3, 4, 5, 6, respectively. Figure 9 shows cut out tiles for images generated for zoom levels 0, 1, and 2. By converting a complex network into multiple tile images and only displaying tiles surrounding the mouse pointer, users can navigate and zoom in & out extremely fast, and the performance is independent of the number of nodes and edges found in the MMP network.  Each node in the MMP network represents a ChEMBL compound and is colored according to its intended target class.  When the node is double clicked, the structure of the ChEMBL compound is retrieved, and the molecule viewer app node displays the structure.  This again could be used as a starting point to link other app nodes (Figure 10).

**CONCLUSION**

We use analytics and productive software whether standalone or browser-based to solve specific tasks, and software programs have evolved to do that job very well.  As data that is needed by those programs move from user's desktop to the cloud and internet connections are getting faster, software developers have redesigned their programs to incorporate advantages offered by the change.  The current trend is to port a standalone program to a browser-based application to better utilize data stored in the cloud, to simplify the installation process, and to improve their revenue stream.

The real opportunity, however, lies within relationships found among diverse data and applications which use them.  As more data are added and their relationships uncovered, it will be easier to exploit that relationship.  Many applications make use of data relationships using relational database management systems, and low-level sharing of software components between applications can be achieved using component object model based technologies, which led to MicroSoft's Object Linking and Embedding and Active X.[39]  Interestingly, existing technologies tend to hide relationships from users unless requested, and, often, dedicated visual and analytics software are required to study them.  And no software exists to understand application relationships and how users use them.  Due to the lack of a standard and a long history of data-centric and task oriented software development, it would be difficult to change the current trend.

The connected applications concept implemented in COMBINE illustrates that many advantages exist if applications are connected at the higher level and application relationships can be visualized. Experience is a sequence of events, and this event in COMBINE is an app node. By stringing together app nodes, COMBINE users can create an experience, a story. A story that can be shared with other users to impart insight.

Table 1.  The list of resources that can be accessed from COMBINE.

| Resources | Description/URL | Content (stats can vary as they are updated regularly) |
|---|---|---|
| ChEMBL[40] | Manually curated drug-like bioactive compounds<br>https://www.ebi.ac.uk/chembl | 11,538 targets, 1,735,442 compounds, 14,675,320 activities, and 67,722 publications (version 23) |
| Open PHACTS[41] API | API to access datasets integrated in the Open PHACTS Discovery Platform<br>https://www.openphacts.org | API to access data found in ChEBI, ChEMBL, SureChEMBL, ChemSpider, ConceptWiki, DisGeNET, DrugBank, Gene Ontology, neXtProt, UniProt and WikiPathways |
| SureChEMBL[42] | Chemically annotated patent document database<br>https://www.surechembl.org | Open patent data containing over 17 million compounds |
| UniChem[43] | Unified chemical structure cross-referencing and identifier tracking system<br>https://www.ebi.ac.uk/unichem | Cross-references of over 151 million structures from 33 data sources |
| UniProt[44] | Protein knowledgebase<br>http://www.uniprot.org | protein sequence and functional information of reviewed (555,100) and unreviewed (88,032,926) molecular targets |
| PDB[45] | Protein data bank<br>https://www.rcsb.org | 41,817 distinct protein sequences, 37,068 structures of human sequences, and 9,503 nucleic acid containing structures |
| BioDBnet[46] | Biological database network<br>https://biodbnet-abcc.ncifcrf.gov | 207 distinct nodes and 738 edges |
| BioGRID[47] | Interaction datasets<br>https://thebiogrid.org | 63,354 publications, 1,493,749 protein and genetic interactions, 27,785 chemical associations, and 38,559 post translational modifications |
| LINCS[48] | The Library of Integrated Network-based Cellular Signatures (LINCS)<br>http://lincsportal.ccs.miami.edu | 350 datasets, 41,847 small molecules 1,127 cells, 978 genes, 1,469 proteins, 155 peptide, and 8 antibodies |
| ZINC[49] | Purchasable compound database<br>http://zinc15.docking.org | Over 100 million compounds in 3D formats |
| TCIA[50] | The cancer imaging archive<br>http://www.cancerimagingarchive.net | 72 collections, 36 cancer types, and 34,959 subjects |
| PepBank[51] | Peptide database<br>http://pepbank.mgh.harvard.edu | 21,691 peptides |

| Metrabase[52] | Metabolism and transport database http://www-metrabase.ch.cam.ac.uk | 20 transporters and 13 CYPs, 3,438 compounds, 11,649 interaction records, and 1,211 literature references |
|---|---|---|
| Firebrowse[53] | Cancer data exploration tool http://firebrowse.org | 28 cohorts and 14,729 cases |
| Human kinome[54] | Aligned human kinases http://kinase.com/human/kinome/groups/ePK.aln | 491 sequences |

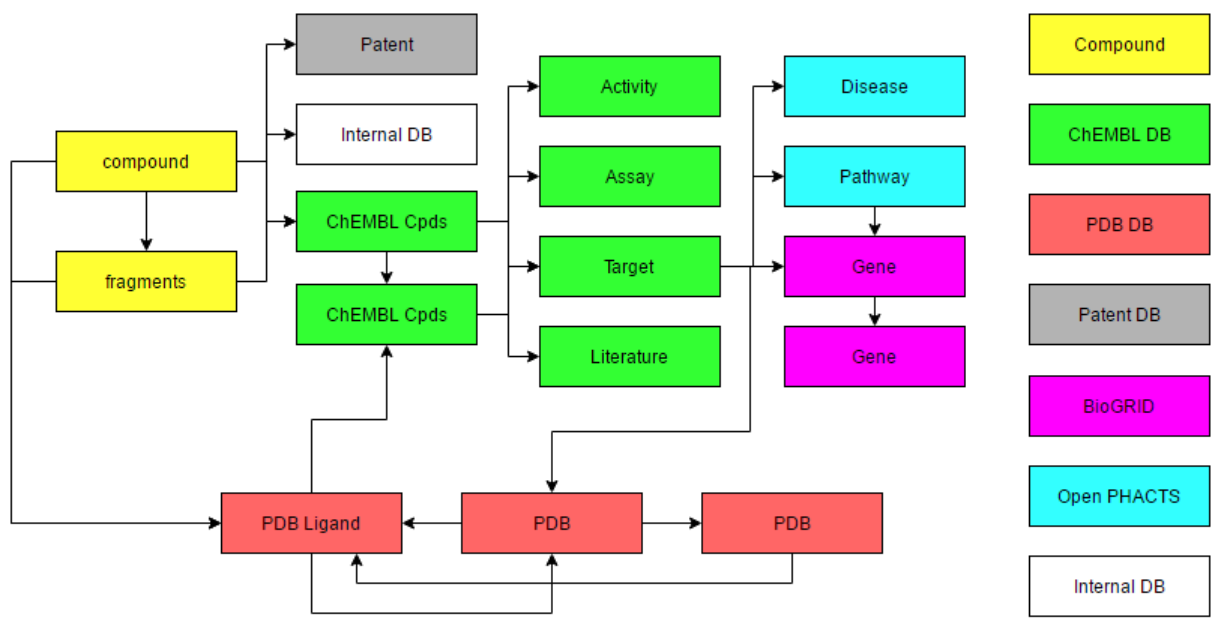

(a)

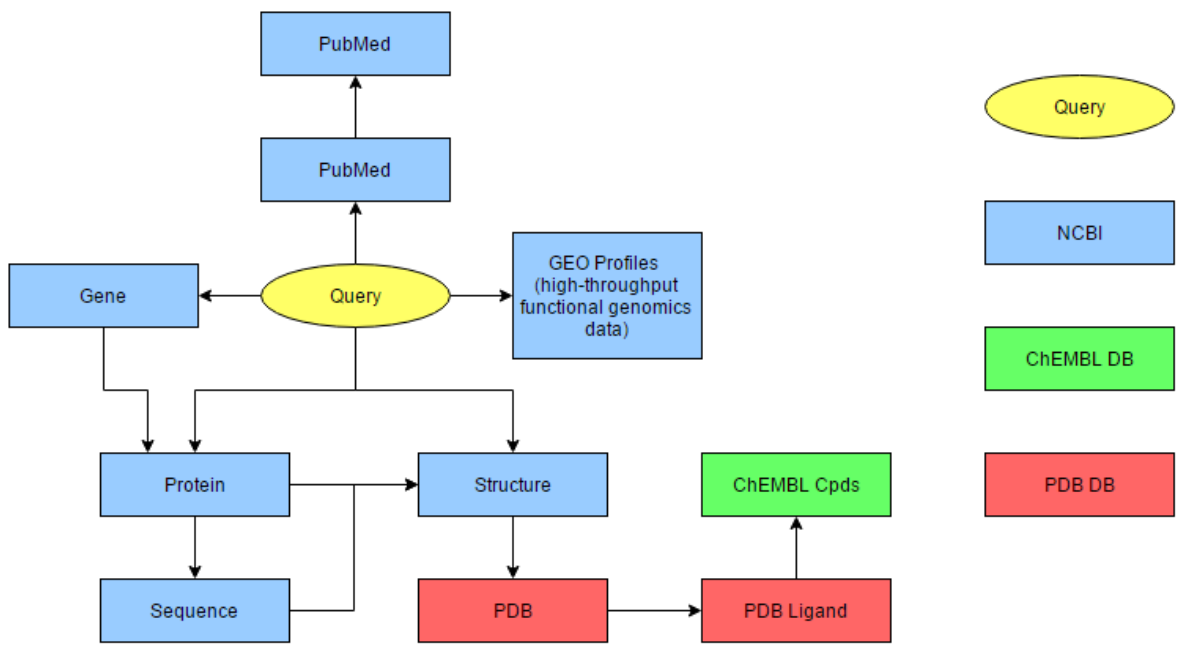

(b)

Figure 1. Two example workflows described as flowcharts. Colors represent databases. (a) A workflow describing how to find the biological activity of a compound is shown. (b) A workflow describing how to identify a small molecule related to a query is shown.

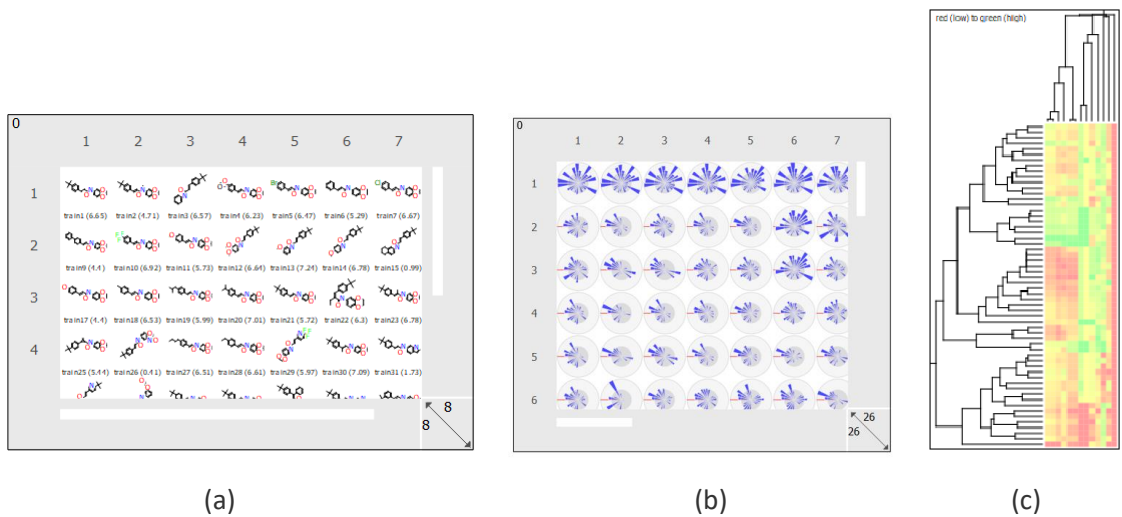

(a)            (b)            (c)

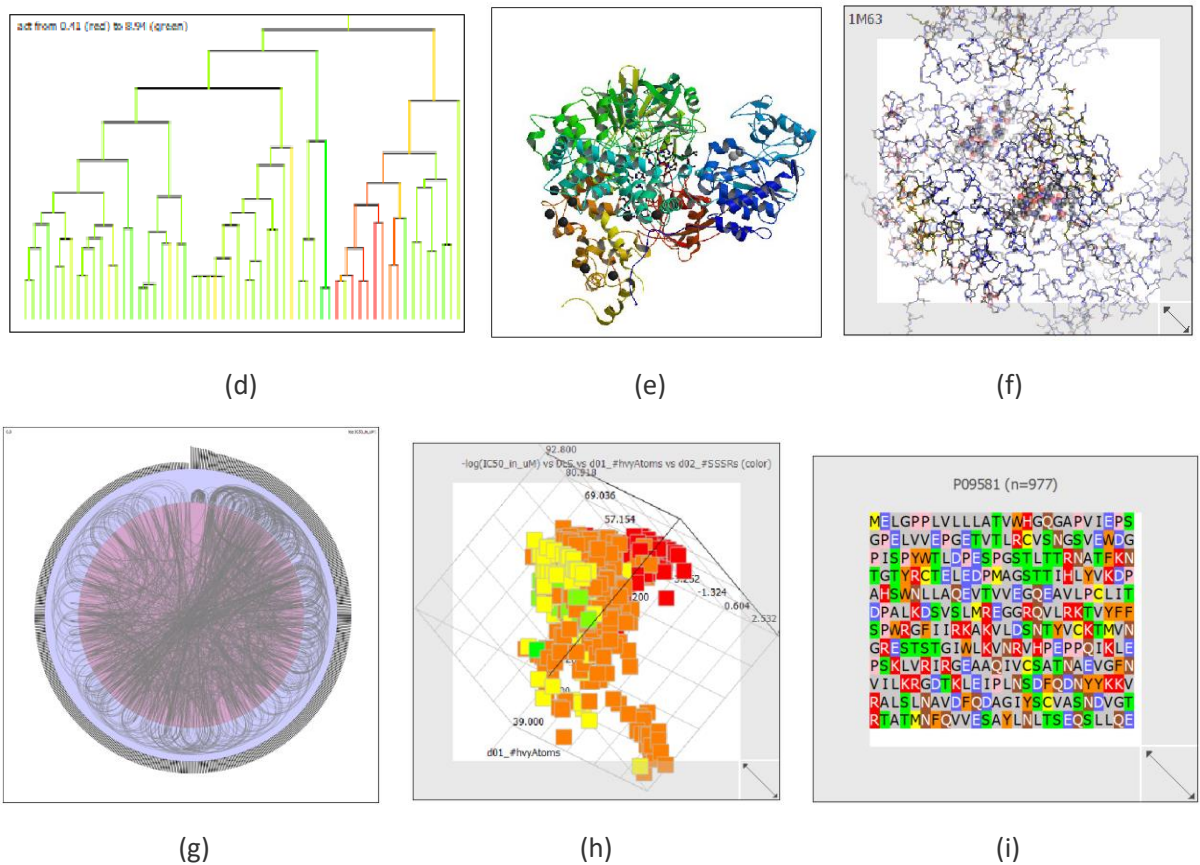

Figure 2. Example app nodes found in COMBINE. (a) An app node containing 58 compounds is shown. The number at the top left corner represents the current zoom level. The row and column numbers are also shown. (b) 35 physicochemical properties are represented as radial, pie shaped, bar charts. (c) A heatmap generated using 58 structures and 13 properties is shown. Colors range from red (low) to yellow (middle) to green (high). (d) A hierarchical clustering result is visualized using a dendrogram. (e) The image of 1M63 is displayed. (f) An interactive molecule viewer displaying the structure of 1M63 is shown. (g) A chord diagram generated using 673 compounds. Curved lines link compounds that have ≥ 0.8 similarity. Straight lines surrounding the circle represent activities. (h) An interactive 3D scatter plot generated using 673 compounds is shown. (g) A protein sequence viewer is used to display the sequence of P09581.

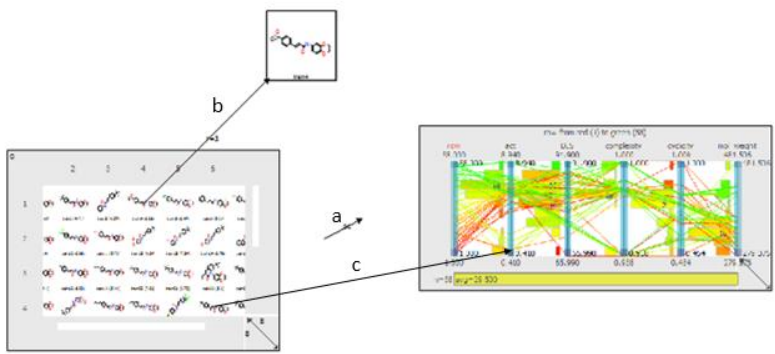

Figure 3. An example of a knowledge network consisting of three app nodes, a structure table, a structure viewer, and parallel coordinates. Edges describe relationships between app nodes: (a) between app nodes, (b) between a data point in an app node to another app node, and (c) between a data point in an app node to a data point in another app node.

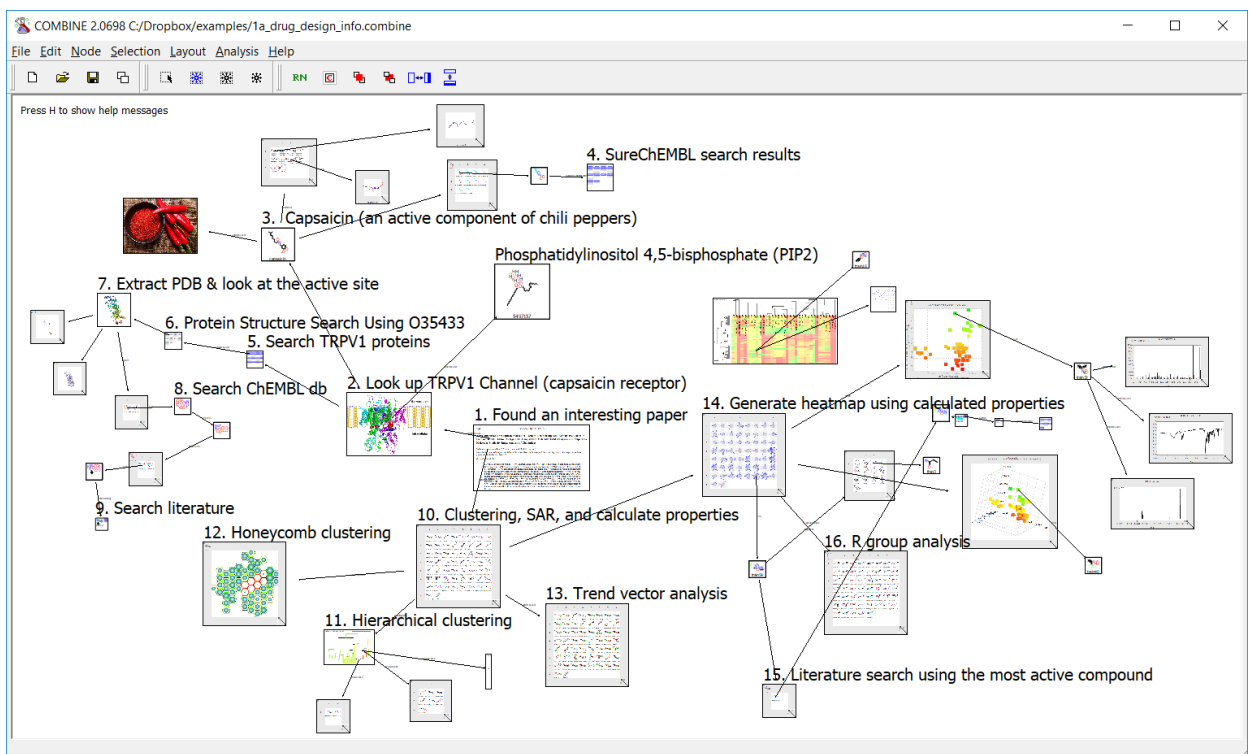

(a)

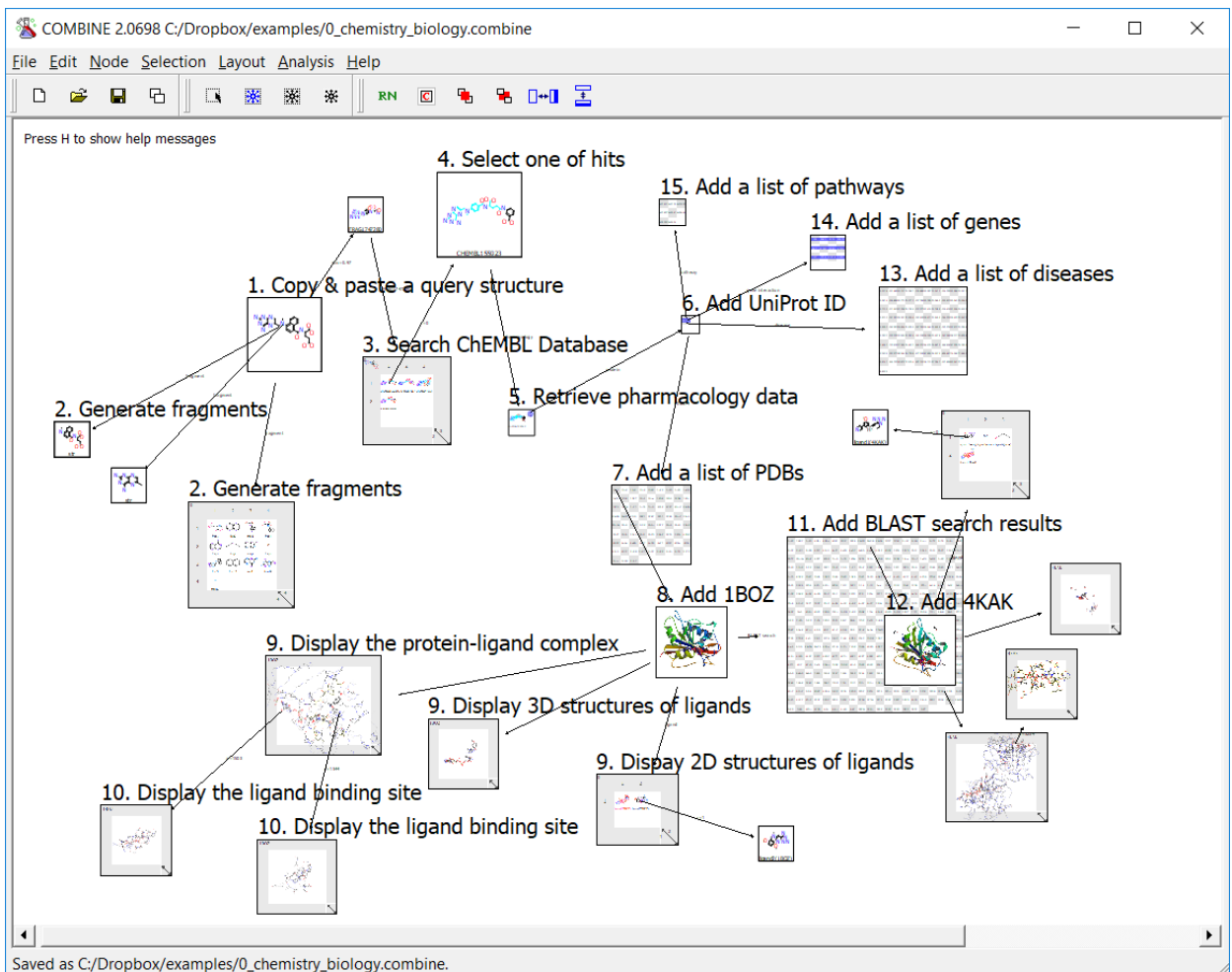

(b)

Figure 4. Example knowledge networks describing data organization and data sharing: (a) annotating a journal of medicinal chemistry paper and (b) linking chemical and biological data.

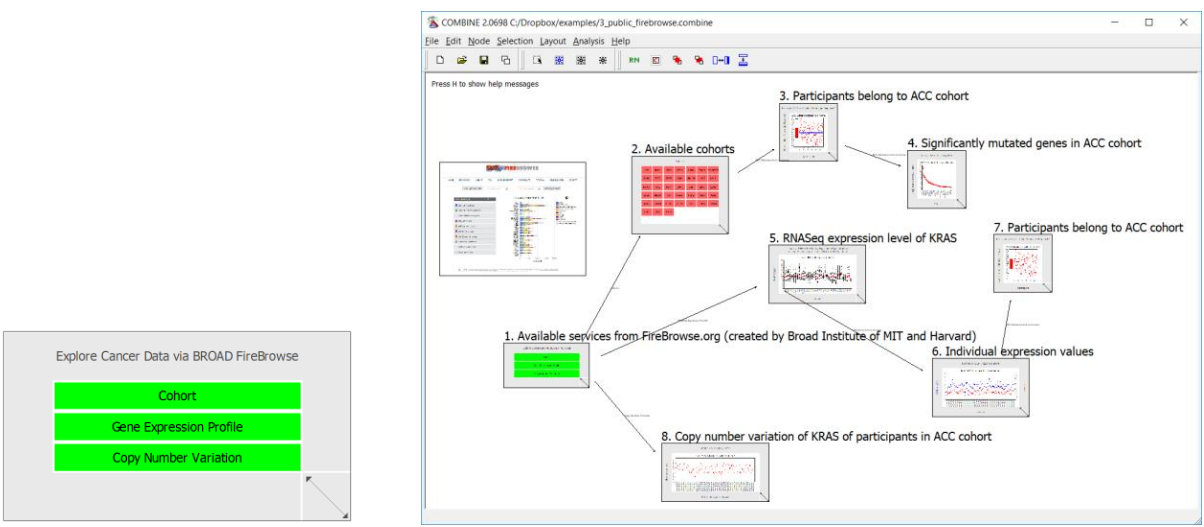

(a)                                                                                          (b)
Figure 5.  Three button UI created using Firebrowse's Web APIs.  (a) The UI and (b) a knowledge network generated are shown.

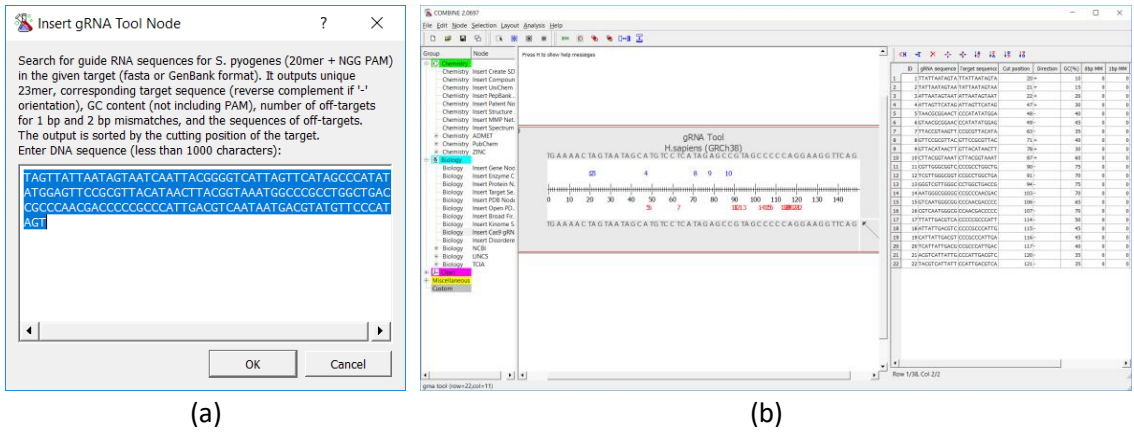

(a)                                                                                          (b)
Figure 6.  gRNA tool app node.  (a) The UI and (b) an app node displaying the output are shown.

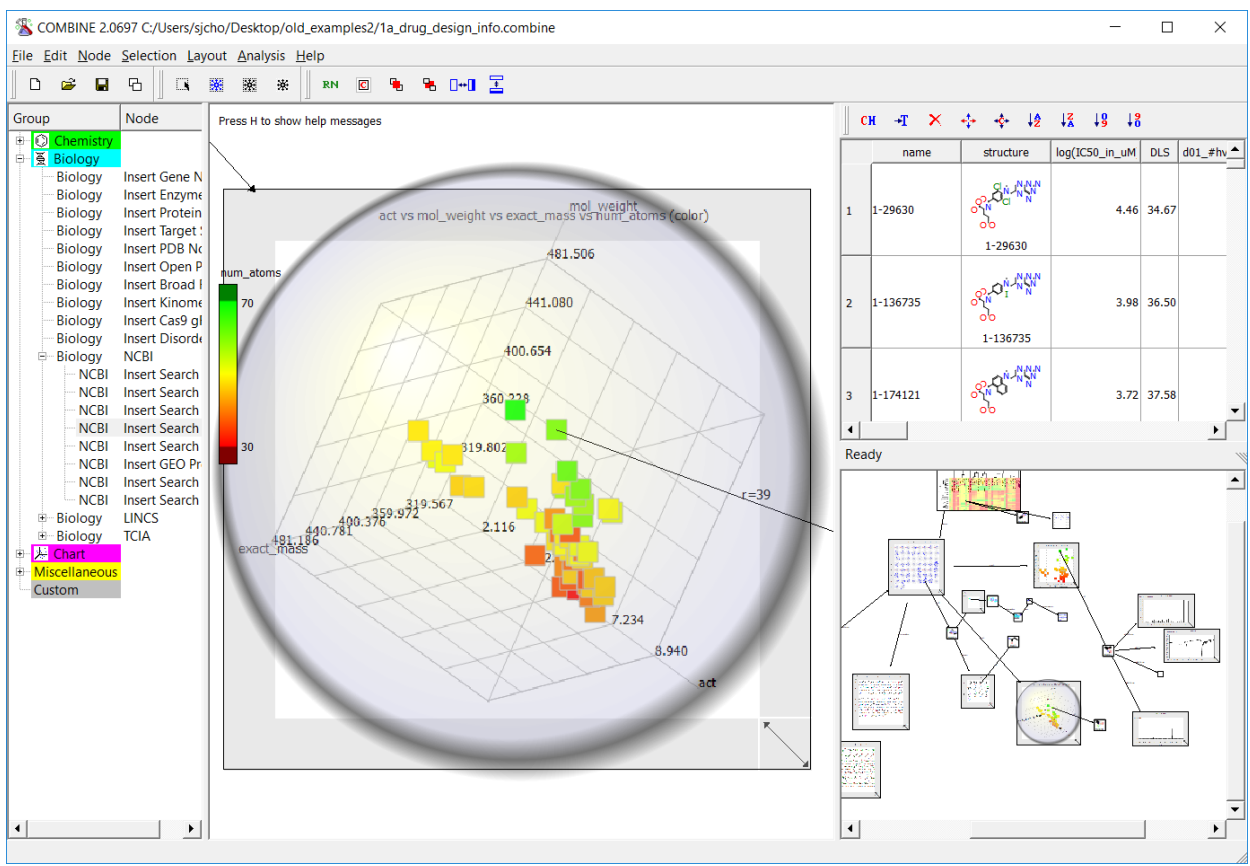

Figure 7. The preview panel is located at the bottom right. The image of a lens is used to indicate the currently focused area.

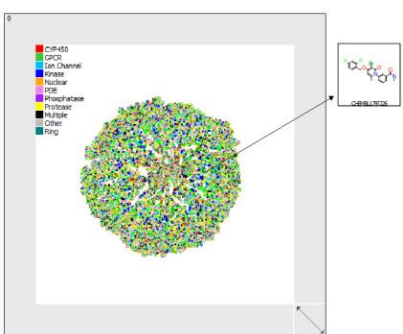

(a)

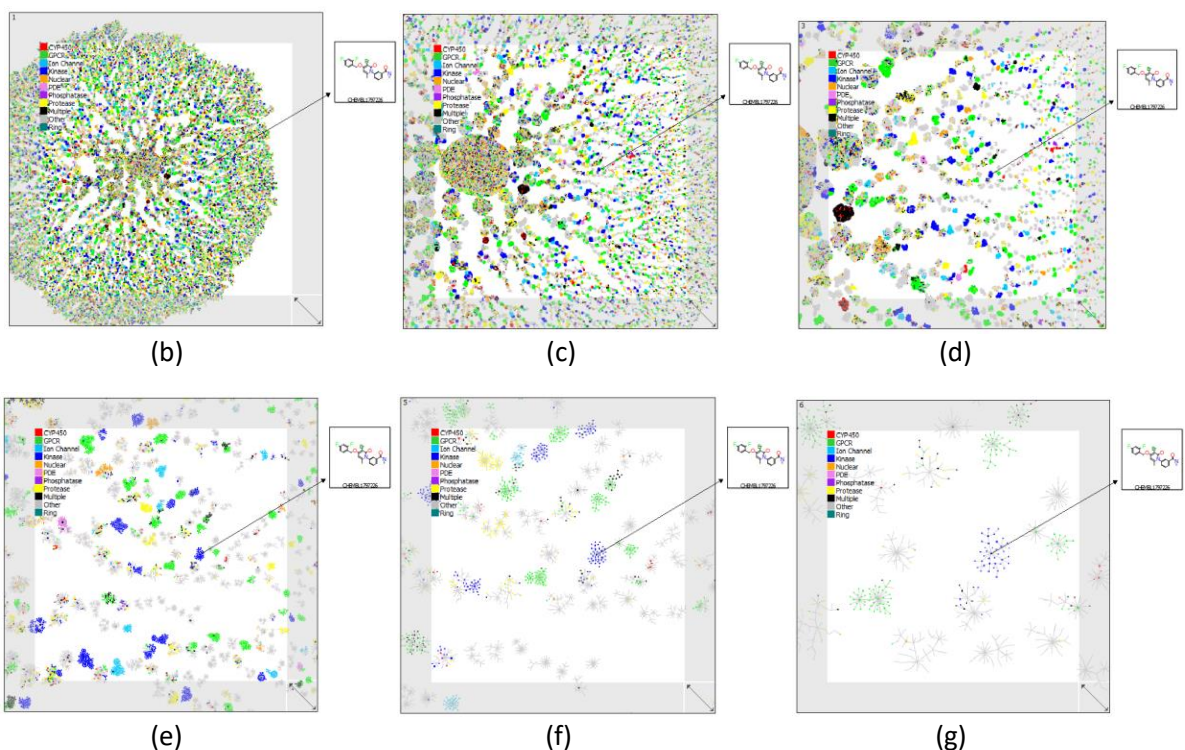

Figure 8. **An example of the** app node level details-on-demand visualization. A matched molecular pair network created using 782,524 compounds with 14,680,477 edges is shown. The top level (a) and zoomed levels 1-6 (b-g) are shown. The mouse scroll bar is used to control the zoom level.

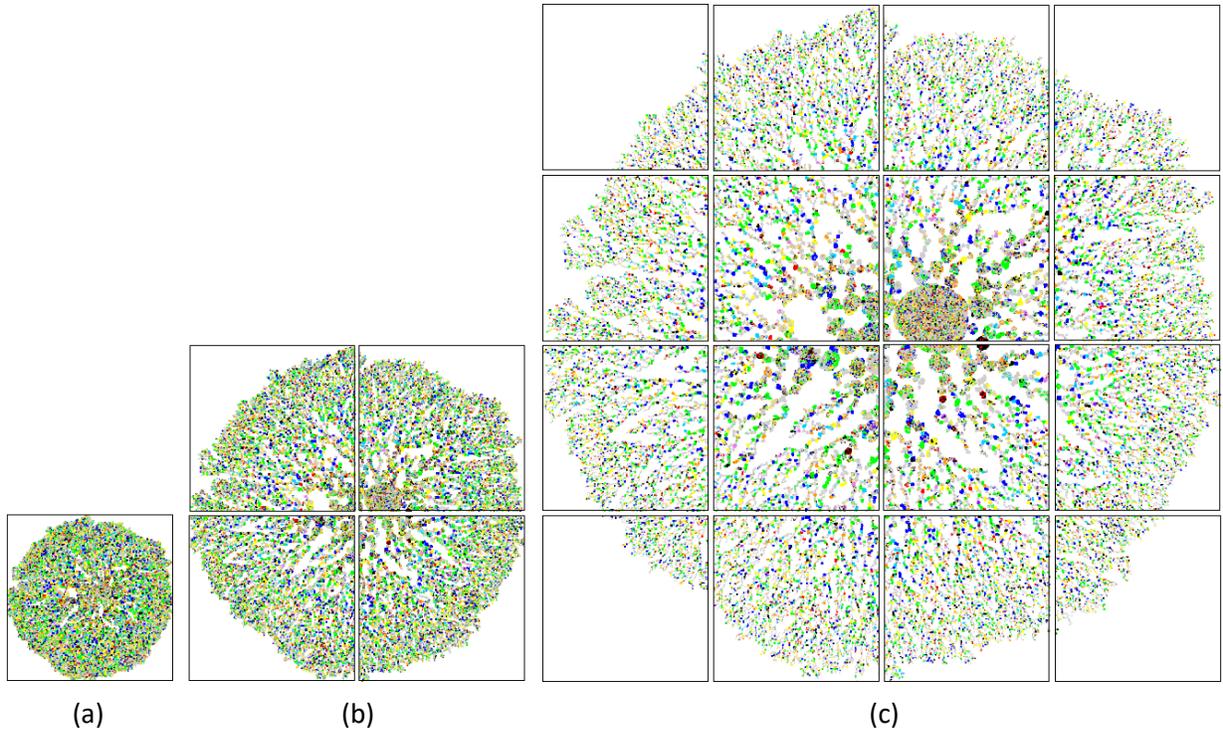

 (a) (b) (c)

Figure 9. Tiles generated after cutting (a) 256x256, (b) 512x512, and (c) 1024x1024 pixel images representing zoom levels 0, 1, and 2 respectively.

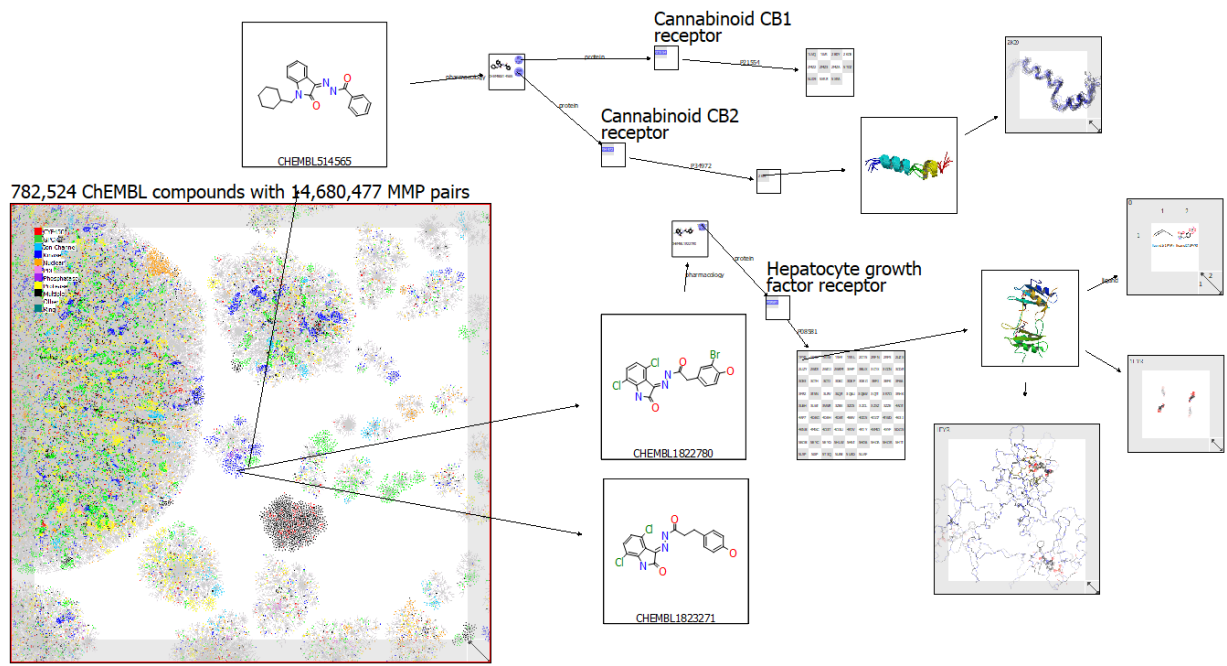

Figure 10. The MMP network viewer is used as a starting app node to generate this knowledge network. After displaying the structures of three ChEMBL compounds, the pharmacology data of two ChEMBL compounds were added.